# The Invisible COVID-19 Crisis: Post-Traumatic Stress Disorder Risk Among Frontline Physicians Treating COVID-19 Patients

**Sayanti Mukherjee, PhD** (corresponding author)[1]
Department of Industrial and Systems Engineering, School of Engineering and Applied Sciences
University at Buffalo – The State University of New York, Buffalo NY
sayantim@buffalo.edu

**Lance Rintamaki, PhD**
Department of Communication, College of Arts and Sciences
University at Buffalo – The State University of New York, Buffalo NY
rlance@buffalo.edu

**Janet L. Shucard, PhD**
Department of Neurology, Jacobs School of Medicine and Biomedical Sciences
University at Buffalo – The State University of New York, Buffalo, NY
shucard@buffalo.edu

**Zhiyuan Wei, MS**
Department of Industrial and Systems Engineering, School of Engineering and Applied Sciences
University at Buffalo – The State University of New York, Buffalo, NY
zwei7@buffalo.edu

**Lindsey E. Carlasare, MBA**
Health Care Research and Policy Analysis
American Medical Association, Chicago, IL
Lindsey.Carlasare@ama-assn.org

**Christine A. Sinsky, MD, FACP**
Professional Satisfaction
American Medical Association, Chicago, IL
Christine.Sinsky@ama-assn.org

[1] 411 Bell Hall, Department of Industrial and Systems Engineering, University at Buffalo (SUNY), North Campus, Buffalo NY 14260; Ph: 716-645-4669

## Abstract

This study evaluated posttraumatic stress disorder (PTSD) among frontline US physicians (treating COVID-19 patients) in comparison with second-line physicians (not treating COVID-19 patients), and identified the significance and patterns of factors associated with higher PTSD risk. A cross-sectional, web-based survey was deployed during August and September, 2020, to practicing physicians in the 18 states with the largest COVID-19 cases. Among 1,478 responding physicians, 1,017 completed the PTSD Checklist (PCL-5). First, the PCL-5 was used to compare symptom endorsement between the two physician groups. A greater percentage of frontline than second-line physicians had clinically significant endorsement of PCL-5 symptoms and higher PCL-5 scores. Second, logistic regression and seven nonlinear machine learning (ML) algorithms were leveraged to identify potential predictors of PTSD risk by analyzing variable importance and partial dependence plots. Predictors of PTSD risk included cognitive/psychological measures, occupational characteristics, work experiences, social support, demographics, and workplace characteristics. The ML algorithm, random forest, outperformed all other models, indicating presence of complex nonlinearity in the predictor—PTSD risk relationships. Importantly, this model identified patterns of both damaging and protective predictors of PTSD risk among frontline physicians. Key damaging factors included depression, burnout, negative coping, fears of contracting/transmitting COVID-19, perceived stigma, and insufficient resources to treat COVID-19 patients. Protective factors included resilience and support from employers/friends/family/significant others. This study underscores the value of ML algorithms to uncover nonlinear relationships among protective/damaging risk factors for PTSD in frontline physicians, which may better inform interventions to prepare healthcare systems for future epidemics/pandemics.

## Highlights:

- Physicians working at the frontlines of COVID-19 are at high risk for PTSD
- Depression, burnout and negative coping strategies are top ranked PTSD risk factors
- Fears of contracting/transmitting COVID-19 and perceived stigma increase PTSD risk
- Higher resilience and support from employer, friends, and family decrease PTSD risk
- Nonlinear relationships exist between PTSD risk and specific risk/protective factors

## 1. Introduction

During previous epidemics, including SARS-2003, H1N1 Influenza-2009, and MERS-2012, frontline physicians (i.e., those treating infected patients) endured formidable social, psychological, and emotional stressors (De Brier et al., 2020; Goulia et al., 2010; Imai et al., 2010; Khalid et al., 2016); yet, COVID-19 far exceeds in scope and scale the devastation wrought by these earlier outbreaks. Physicians have been overrun by caseloads of acutely ill patients (Restauri and Sheridan, 2020; Wu et al., 2009), insufficient resources (Marco et al., 2020; Restauri and Sheridan, 2020), and risks inherent to working with a new and highly infectious disease, culminating in the deaths of over 596,572 Americans (Jun 11, 2021) (Centers for Disease Control and Prevention (CDC), 2021). A recent survey found that 30% of physicians experienced high stress, anxiety, or depression due to COVID-19's impact, and 62% had considerable fear of contracting or transmitting COVID-19 (Linzer et al., 2021). The traumatic nature of COVID-19 associated stressors makes post-traumatic stress disorder (PTSD) of special consequence for COVID-19 frontline physicians (Carmassi et al., 2020).

As defined by the Diagnostic and Statistical Manual of Mental Disorders (DSM-5), PTSD is a psychiatric disorder that follows exposure to a traumatic event (Criterion A), and is characterized by four behavioral, cognitive, and emotional symptom criteria: (A) intrusive, distressing thoughts; (B) persistent avoidance of trauma-related stimuli; (C) alterations in cognitions and mood; and (D) heightened arousal and reactivity (American Psychiatric Association, 2013). These symptoms can persist for decades (Rintamaki et al., 2009). Physicians with PTSD are susceptible to negative coping strategies (e.g., substance abuse) in attempt to manage their symptoms, other mental health conditions (e.g., depression), and suicidality (Gradus et al., 2010; Gregory et al., 2019; Liebschutz et al., 2007). Thus, although physicians on the frontlines of emergent infectious diseases are at high risk of PTSD (Chong et al., 2004; Maunder et al., 2003; Nickell et al., 2004; Wu et al., 2009); assessment of these phenomena in the context of COVID-19 is just emerging in the United States (Blekas et al., 2020; Kang et al., 2020; Lai et al., 2020; Li et al., 2020), which leads the world in COVID-19 cases.

The aims of the current research were to: 1) evaluate the symptoms of PTSD among frontline physicians compared to second-line physicians; 2) predict PTSD risk among frontline physicians using nonparametric machine learning algorithms; 3) identify and rank key predictors associated with PTSD risk of frontline physicians; and 4) determine the linear/nonlinear patterns of these predictors. Novel COVID-19 related social, emotional, and cognitive factors were assessed while accounting for other known factors influencing PTSD, which included psychological resilience (Colville et al., 2017; Hamid and Musa, 2017; Winkel et al., 2019), exposure rate (Jung et al., 2020; Kang et al., 2020; Lai et al., 2020), occupational role (Lai et al., 2020; Maunder et al., 2004), age (Sim et al., 2004; Wu et al., 2009), sex (Chong et al., 2004), marital status (Chan and Huak, 2004), isolation (Wu et al., 2009), coping strategies (Chan and Huak, 2004; Maunder et al., 2006; Wu et al., 2009), and social support from family (Chan and Huak, 2004; Su et al., 2007), colleagues and organization (Chan and Huak, 2004; Lancee et al., 2008; Maunder et al., 2006).

## 2. Material and methods

### 2.1. Study Design

Following IRB approval, a cross-sectional, web-based survey developed by our interdisciplinary team was deployed to physicians from the American Medical Association's (AMA) Physician Masterfile database between August 7 and September 26, 2020. Sampling immediately followed the second COVID-19 US contagion peak (Johns Hopkins University & Medicine, 2021). Participation was voluntary and targeted physicians from states reporting the greatest COVID-19 caseloads (more than 40,000 COVID-19 cases as of June 2020) (CDC, 2020), including New York, California, New Jersey, Illinois, Texas, Massachusetts, Florida, Pennsylvania, Michigan, Georgia, Maryland, Virginia, North Carolina, Arizona, Louisiana, Connecticut, Ohio, and Indiana.

**2.2. Participants**

Physicians from all specialties were recruited using the AMA Physician Masterfile, a near-complete record of all US physicians, independent of AMA membership. Canvassing e-mails (including study description and survey link) were sent on August 7 and 26, 2020. From 36,372 physicians (opening invitation), 1,478 responses were recorded, of which 1,017 responses (completing PTSD Checklist: PCL-5) were analyzed (***eFigure 1***).

**2.3. Outcome variable: Post-Traumatic Stress Disorder**

We employed the PTSD Checklist, a commonly used research and clinical screening questionnaire based on the DSM-5 (PCL-5) (Weathers et al., 2013) to assess PTSD symptoms (see Table 2). Using this 20-item, 5-point scale (0=*not at all* to 4=*extremely*), respondents rated how bothered they were by PTSD symptoms in the past month. The total score range is 0–80, with 33 or greater indicating probable PTSD. A diagnosis of PTSD can only be made by a trained clinician using an in-person interview; thus, we use the term “probable PTSD” to indicate physicians with the highest symptom ratings who are likely to have PTSD.

To optimize the categorization of physicians into PTSD groups (low PTSD risk to high PTSD risk) for the predictive analyses, we combined the DSM-5 and PCL-5 scoring criteria, similar to methods suggested by the National Center for PTSD (Lu et al., 2006). This procedure included using only PCL-5 items rated as 2 (moderately) or higher, which constitutes clinically significant symptom endorsement, and then applying this level of endorsement to the number of DSM-5 items required for each of the four criterion: at least one item in Criterion B (re-experiencing), one in Criterion C (avoidance), two in Criterion D (negative beliefs), and two in Criterion E (hyperarousal) (American Psychiatric Association, 2013). Table 3 presents the frequency of physicians who endorsed PCL-5 symptoms as 2 or higher. Physicians working directly with COVID-19 patients were designated as *frontline physicians* and those who were not, as *second-line physicians*. For modeling analyses, the frontline physicians then were categorized into two groups: a "high risk group" (probable/subclinical PTSD symptoms), and a "low risk group" (none/pre-subclinical PTSD symptoms) (***eMethods 1*** details PTSD classification).

### 2.4. Predictor Candidates

The Patient Health Questionnaire (PHQ-9) [9 items, 4-level Likert scale: 0-3 score range per item, total score: 0-27] (Kroenke et al., 2001) and the single-item, 5-point burnout scale [1 item, 5-level Likert scale: 1-5 score range] (Dolan et al., 2015) were used to assess the severity of symptoms of **depression** and **burnout,** respectively. We categorized depression into five levels based on total PHQ-9 scores (Kroenke et al., 2001): *minimal* [total score=1-4]; *mild* [total score=5-9]; *moderate* [total score=10-14]; *moderately severe* [total score=15-19]; and *severe* [total score=20-27]. **Resilience** and **stress coping characteristics** were measured, respectively, with the Connor-Davidson Resilience Scale (CD-RISC-10) [10 items, 5-level Likert scale: 0-4 score range per item, total score: 0-40] (Davidson, 2020) and, the Brief-COPE Scale [28-item, 4-level Likert scale: 1-4 score range per item] (Carver, 1997). The Brief-COPE scores indicate individuals' negative/positive dominant coping strategies among 14 categories, each scored separately with a range of 2-8. The 14 categories include self-distraction, active coping, denial, substance abuse, use of emotional support, use of instrumental support, behavioral disengagement, venting, positive reframing, planning, humor, acceptance, religion, and self-blame (Carver, 1997).

**Occupational characteristics** and **COVID-19 specific experiences** included living arrangement changes, workload, non-routine work, resource availability, decision-making, exposure rates (e.g., time with COVID-19 patients, intubation/aerosol-generating procedures of suspected/confirmed COVID-19 patients), perceived stigma from treating COVID-19 patients, and turnover intent (switching units/teams, leaving current employer, or leaving healthcare entirely). **Perceived organizational and social support** was assessed using the 8-item 7-point Survey of Perceived Organizational Support scale [8 items, 7-level Likert scale: 1-7 score range per item] (SPOS; items 1, 3, 7, 9, 17, 21, 23, 27) (Eisenberger et al., 1986), which was classified into positive support (item numbers: 1, 9, 21, 27) (Eisenberger et al., 1986) and negative support (item numbers: 3, 7, 17, 23) (Eisenberger et al., 1986) for our analysis. Perceived available support from family, friends, and significant others was measured employing the 3-item Multidimensional Scale of Perceived Social Support (MSPSS) (Zimet et al., 1988).

**Demographics** included age, sex, ethnicity, race, immigration status, and marital status. **Workplace characteristics** included training/years of experience, primary work setting, hospital type, and work setting within hospital.

## 2.5. Statistical Analysis

Independent samples $t$-tests and Chi-square ($\chi^2$test) analyses were used to test the PCL-5 scores (Table 2) and frequency of endorsed symptoms (Table 3) between frontline and second-line physicians, respectively. Chi-square analyses ($\chi^2$ test) and $t$-tests were also implemented after obtaining the key variables from the model implementation to determine if the findings were considered significant at 2-tailed $P \leq .05$.

If complex, nonlinear interplay exists for PTSD risk factors, then conventional, linear models (e.g., logistic regression) would likely fail to capture this phenomenon. Therefore, we trained and tested seven machine learning models alongside logistic regression, including random forest, bootstrap-aggregating (bagging), Naïve Bayes, gradient boosting method, Bayesian additive regression trees, support vector machines, and neural networks. Selection of an optimal predictive model was based on the generalization performance of the models (Hastie et al., 2008; James et al., 2013) (details in ***eMethods 2).***

Next, *to identify the key factors of PTSD risk*, we conducted variable importance (VI) analysis, where importance of each PTSD predictor variable was measured using the Gini index (Breiman, 2001). We implemented partial dependency analysis to evaluate the marginal effect of a particular predictor on PTSD risk, while keeping other variables constant (Breiman, 2001; Greenwell, 2017). Quantified marginal effects could indicate an increase/decrease in PTSD risk with increase/decrease in magnitude of the predictor, thus helping to categorize predictors as damaging/protective factors, respectively. All analyses were performed in R (version-3.1) and RStudio (version-1.1.463).

# 3. Results

### 3.1. Participants' Socio-demographic Characteristics

**Table 1** presents demographic data and statistical comparisons between frontline and second-line physicians. The groups differed across age, years in practice, sex, primary work setting, current work status, and underlying conditions. Frontline physicians were younger and less experienced than second-line physicians. Sex composition of frontline physicians was similar, whereas second-line skewed female. Most frontline physicians worked in hospitals, group-practice, or academic medical centers, whereas second-line physicians worked in group-practice, academic medical centers, or solo-practice. Lastly, more frontline than second-line physicians were employed full-time, whereas more second-line than frontline physicians worked part-time.

**Table 1.** Demographics of the sample population (^ = $t$-test; * = $\chi^2$ test)

| Sample Demographic Characteristics | Frontline Physicians: No. (%) (n= 717) | Second-line Physicians: No. (%) (n= 300) | Significance test (*P* value) |
|---|---|---|---|
| **Age, mean (SD)** | 51.35 (11.3) | 54.57 (11.81) | 0.0283^ |
| Missing | 160 | 53 | |
| **Sex** | | | |
| Male | 317 (44.21) | 98 (32.67) | |
| Female | 323 (45.05) | 181 (60.33) | |
| Non-conforming, non-binary, transgender | 1 (0.14) | 0 (0) | < .001* |
| Prefer not to answer | 6 (0.84) | 1 (0.33) | |
| Missing | 70 | 20 | |
| **Ethnicity** | | | |
| Hispanic / Latino | 59 (8.23) | 22 (7.33) | |
| Non-Hispanic | 589 (82.15) | 258 (86.0) | 0.623* |
| Missing | 69 | 20 | |
| **Race** | | | |
| American Indian or Alaskan Native | 1 (0.14) | 0 (0) | |
| Asian | 91 (12.69) | 26 (8.67) | |
| Black or African American | 25 (3.49) | 14 (4.67) | |
| Native Hawaiian or Pacific Islander | 1 (0.14) | 0 (0) | 0.166* |
| White | 466 (64.99) | 222 (74.0) | |
| Others | 55 (7.67) | 17 (5.67) | |
| Missing | 78 | 21 | |
| **Immigration Status** | | | |
| U.S. immigrant | 109 (15.20) | 40 (13.33) | |
| Not an U.S. immigrant | 540 (75.31) | 240 (80.0) | 0.390* |
| Missing | 68 | 20 | |

| Sample Demographic Characteristics | Frontline Physicians: No. (%) (n= 717) | Second-line Physicians: No. (%) (n= 300) | Significance test (*P* value) |
|---|---|---|---|
| **Relationships Status** | | | |
| Single | 94 (13.11) | 30 (10.0) | |
| Married | 512 (71.41) | 230 (76.67) | |
| Partnered | 38 (5.3) | 12 (4.0) | 0.096* |
| Widow / Widower | 5 (0.7) | 6 (2.0) | |
| Missing | 68 | 22 | |
| **Number of years practicing, mean (SD)** | 21.76 (10.94) | 25.49 (12.57) | 0.003^ |
| Missing | 81 | 22 | |
| **Primary Work Setting** | | | |
| Academic medical center | 146 (20.36) | 57 (19.0) | |
| Group practice | 150 (20.92) | 82 (27.33) | |
| Hospital | 182 (25.38) | 24 (8.0) | |
| Solo practice | 55 (7.67) | 40 (13.33) | < .001* |
| Two-physician practice | 24 (3.35) | 18 (6.0) | |
| Outpatient center | 42 (5.86) | 27 (9.0) | |
| Others | 41 (5.72) | 30 (10.0) | |
| Missing | 77 | 22 | |
| **Current Working Status** | | | |
| Full-time | 567 (79.08) | 209 (69.67) | |
| Part-time | 62 (8.65) | 59 (19.67) | |
| Furloughed | 5 (0.7) | 1 (0.33) | < .001* |
| Laid off | 1 (0.14) | 3 (1.0) | |
| On leave | 7 (0.98) | 6 (2.0) | |
| Missing | 75 | 22 | |
| **Having underlying health conditions** | | | |
| Yes | 242 (33.75) | 124 (41.33) | |
| No | 452 (63.04) | 162 (54.0) | 0.024* |
| Not sure | 23 (3.21) | 14 (4.67) | |
| **Pregnancy (self or partner)** | | | |
| Yes, pregnant or with a newborn | 50 (6.97) | 14 (4.67) | 0.215* |
| No | 667 (93.03) | 286 (95.3) | |

## 3.2. Comparison of PTSD Symptoms Between Physician Groups

Overall, 717 frontline physicians and 300 second-line physicians completed the PCL-5. **Table 2** presents the PCL-5 data using the full-scale scores of 0 to 4. Section 1 lists the means (SDs) of frontline vs. second-line physicians for each of the 20 PCL-5 items. More frontline than second-line physicians had significantly higher scores for all of the items in Criterion B (re-experiencing) and C (avoidance), four of seven items in Criterion D (negative cognition and mood), and four of six items in Criterion E (heightened arousal). Section 2 of Table 2 lists PCL-5 composite criterion scores and the PCL-5 total score (calculated by summing all PCL-5 item scores) for both groups. Frontline compared to second-line physicians had significantly higher criterion scores and PCL-5 total score.

**Table 2.** PTSD Checklist (PCL-5) mean score and standard deviation differences between frontline and second-line physicians

| PCL-5 items for each DSM-5 Symptom cluster (Criterion B, C, D, E) | Frontline physicians (n= 717) | Second-line physicians (n= 300) | Significance (*P value)* |
|---|---|---|---|
| | **Mean (SD)** | **Mean (SD)** | ***t*-test** |
| **Section 1** | | | |
| **Criterion B: Re-experiencing Symptoms** | | | |
| 1. Disturbing memories | 0.82 (0.98) | 0.49 (0.82) | < .001 |
| 2. Disturbing dreams | 0.59 (0.91) | 0.35 (0.75) | < .001 |
| 3. Flashbacks | 0.35 (0.80) | 0.17 (0.53) | < .001 |
| 4. Feeling upset when reminded of event | 0.72 (0.97) | 0.50 (0.86) | < .001 |
| 5. Physical reactions when reminded of event | 0.47 (0.86) | 0.33 (0.75) | 0.009 |
| **Criterion C: Avoidance Symptoms** | | | |
| 6. Avoiding memories related to event | 0.64 (0.95) | 0.44 (0.85) | 0.001 |
| 7. Avoiding external reminders of event | 0.53 (0.90) | 0.37 (0.78) | 0.003 |
| **Criterion D: Negative alterations in cognitions, mood** | | | |
| 8. Trouble remembering part of event | 0.35 (0.77) | 0.16 (0.59) | < .001 |
| 9. Negative beliefs about self, other, world | 0.60 (1.03) | 0.49 (0.87) | 0.1 |
| 10. Blaming yourself | 0.57 (0.94) | 0.40 (0.83) | 0.005 |
| 11. Feeling fear, horror, anger, guilt, shame | 0.66 (1.00) | 0.55 (0.88) | 0.07 |
| 12. Loss of interest in activities | 0.78 (1.06) | 0.60 (0.86) | 0.006 |
| 13. Feeling distant from others | 1.23 (1.21) | 1.09 (1.16) | 0.086 |
| 14. Trouble feeling positive feelings | 0.76 (1.05) | 0.58 (0.91) | 0.005 |
| **Criterion E: Heightened arousal and reactivity** | | | |
| 15. Irritable behavior | 0.89 (1.01) | 0.69 (0.88) | 0.002 |
| 16. Taking risks | 0.25 (0.65) | 0.12 (0.39) | < .001 |
| 17. Hypervigilance: super alert, on guard | 0.88 (1.11) | 0.69 (0.97) | 0.006 |
| 18. Feeling jumpy, or easily startled | 0.53 (0.94) | 0.41 (0.77) | 0.028 |
| 19. Difficulty concentrating | 0.83 (1.04) | 0.74 (0.91) | 0.195 |
| 20. Trouble falling or staying asleep | 1.16 (1.19) | 1.03 (1.12) | 0.1 |
| **Section 2** | | | |
| **PCL-5 Criterion and Total Scores (range)** | | | |
| Criterion B: Total score (0-20) | 2.95 (3.87) | 1.83 (3.11) | < .001 |
| Criterion C: Total score (0-8) | 1.17 (1.76) | 0.81 (1.56) | 0.001 |
| Criterion D: Total score (0-28) | 4.94 (5.67) | 3.88 (4.69) | 0.002 |
| Criterion E: Total score (0-24) | 4.54 (4.53) | 3.68 (3.7) | 0.002 |
| Total PCL-5 Score (0-80) | 13.61 (14.39) | 10.2 (11.39) | < .001 |

**Table 3** lists the number (percentage) of frontline vs. second-line physicians who endorsed each of the PCL-5 items with scores of 2 or higher (considered clinically significant). Chi-square ($\chi^2$ test) analyses indicated that a greater number of frontline than second-line physicians endorsed four items in Criterion B, one item in Criterion C, three items in Criterion D, and two items in Criterion E.

**Table 3.** PTSD Checklist (PCL-5) frequency and percentage differences between frontline and second-line physicians

| PCL-5 items for each DSM-5 Symptom cluster (Criterion B, C, D, E) | Frontline physicians (n= 717) | Second-line physicians (n= 300) | Significance (*P value)* |
|---|---|---|---|
| | **No. (%) who endorsed score ≥2** | **No. (%) who endorsed score ≥2** | $\chi^2$ **test** |
| **Section 1** | | | |
| **Criterion B: Re-experiencing Symptoms** | | | |
| 1. Disturbing memories | 144 (20.08) | 30 (10.0) | < .001 |
| 2. Disturbing dreams | 100 (13.95) | 23 (7.67) | 0.007 |
| 3. Flashbacks | 59 (8.23) | 9 (3.0) | 0.004 |
| 4. Feeling upset when reminded of event | 124 (17.29) | 32 (10.67) | 0.010 |
| 5. Physical reactions when reminded of event | 79 (11.02) | 25 (8.33) | 0.240 |
| **Criterion C: Avoidance Symptoms** | | | |
| 6. Avoiding memories related to event | 110 (15.34) | 27 (9.0) | 0.009 |
| 7. Avoiding external reminders of event | 85 (11.85) | 28 (9.33) | 0.290 |
| **Criterion D: Negative alterations in cognitions, mood** | | | |
| 8. Trouble remembering part of event | 59 (8.23) | 10 (3.33) | 0.007 |
| 9. Negative beliefs about self, other, world | 117 (16.32) | 35 (11.67) | 0.072 |
| 10. Blaming yourself | 98 (13.67) | 30 (10.0) | 0.132 |
| 11. Feeling fear, horror, anger, guilt, shame | 105 (14.64) | 35 (11.67) | 0.247 |
| 12. Loss of interest in activities | 147 (20.5) | 41 (13.67) | 0.013 |
| 13. Feeling distant from others | 265 (36.96) | 95 (31.67) | 0.124 |
| 14. Trouble feeling positive feelings | 150 (20.92) | 41 (13.67) | 0.009 |
| **Criterion E: Heightened arousal and reactivity** | | | |
| 15. Irritable behavior | 154 (21.48) | 50 (16.67) | 0.097 |
| 16. Taking risks | 34 (4.74) | 3 (1.0) | 0.006 |
| 17. Hypervigilance: super alert, on guard | 168 (23.43) | 52 (17.33) | 0.038 |
| 18. Feeling jumpy, or easily startled | 94 (13.11) | 29 (9.67) | 0.153 |
| 19. Difficulty concentrating | 142 (19.8) | 53 (17.67) | 0.482 |
| 20. Trouble falling or staying asleep | 239 (33.33) | 83 (27.67) | 0.090 |

**Table 4** presents categorization of physicians into four PTSD symptom groups, as described in Section 2.3. The frontline physicians were then divided into two groups for subsequent machine learning analyses (see Table 4, *low risk group* and *high risk group*). For additional confirmation of the difference between the *low risk group* and the *high-risk group*, PCL-5 total scores were calculated for each group. The mean [SD] of the total PCL-5 score was 6.05 [5.36] for *low risk group*, and 30.59 [13.81] for *high risk group*. The physicians in *low risk* *group* (n=309) and *high risk group* (n=137) answered all questions for each of the predictor measures (***eFigure 1***). This dataset was then used to train and test the machine learning algorithms to predict and evaluate the factors associated with higher PTSD risk.

**Table 4:** PTSD symptom groups: Frequency (percent) of PCL-5 scores for physicians in each PTSD group.

| PTSD Symptom Severity Groups* | Frontline physicians** (n= 717) | Second-line physicians (n= 300) | *p value* Chi Square |
|---|---|---|---|
| | **No. (%)** | **No. (%)** | |
| No PTSD | 413 (57.6) | 202 (67.33) | .005 |
| Pre-subclinical | 83 (11.58) | 34 (11.33) | .998 |
| Sub-clinical | 144 (20.08) | 49 (16.33) | .192 |
| Probable PTSD | 77 (10.74) | 15 (5.0) | .005 |
| | | | |
| **Low risk group** *** | 496 (69.18) | | |
| **High risk group** | 221 (30.82) | | |

* The four groups were determined by combining scoring methods of the DSM-5 and PCL-5 (see ***eMethods 1***)
****** Only data for Frontline physicians were used in the machine learning analyses (*Low risk group* and *High risk group*).
*** *Low risk group* = No PTSD group and pre-subclinical PTSD group; *High risk group* = Sub-clinical PTSD group and probable PTSD group

## Model selection

As discussed previously, selection of an optimal predictive model was based on generalization performance of the models (Hastie et al., 2008; James et al., 2013). This included assessment of goodness-of-fit (***eTable 2***) and predictive accuracy (***eTable 3)***. We conducted significance tests of F1 scores (***eTable 4***) and Area Under the Curve (AUC) measures (***eTable 5***) across every pairing of predictive models. Ultimately, we selected random forest, given it offered: 1) best goodness-of-fit performance among all the models (***eTable 2***); 2) highest overall predictive accuracy (accuracy=82.52%, 95% CI, 81.16-83.89; recall=93.11%, 95% CI, 91.92-94.31) (***eTable 3, eFigure 2***); 3) best prediction performance determined by F1 score (***eTable 4***); and, 4) more interpretability than its competitors such as the support vector machine and other black-box algorithms (Rustam et al., 2019) (***eTable 5)***.

### 3.3. Key Predictors for Increased Risk of PTSD in Frontline Physicians

The central aim of this study was to determine which variables best predict PTSD risk among frontline physicians and how these variables relate to higher risk of PTSD, or possible protection from PTSD. **Fig. 1** depicts the variable rankings (rank no.1—most important predictor) (see ***eMethods 3*** for details). **Table 5** illustrates statistical significance of the top 20 variables indicated in **Fig. 1**. For the continuous variables, we used the $t$-test whereas for the ordered variables we used the $\chi^2$test for comparing if the factors were significantly different between the *low risk group* and *high risk group* of frontline physicians. Our findings are organized into two categories: 1) damaging variable/risk factors (higher scores associated with higher PTSD risk); and, 2) protective variables (higher scores associated with lower PTSD risk). The relationships were obtained using the partial dependence plots (PDPs) (see ***eMethods 3*** for details). The PDPs of the top 20 variables are illustrated in **Fig. 2**.

**Table 5.** Top 20 predictors of PTSD risk in frontline US physicians treating COVID-19 patients (^ = $t$-test; * = $\chi^2$ test)

| Rank | Variable names | Scale Range | Low risk group (N=309) | High risk group (N=137) | VIP | Statistical analysis | *p*-value |
|---|---|---|---|---|---|---|---|
| | | Min –Max | Mean (SD) | Mean (SD) | | | |
| 1 | PHQ-9 Depression score | 0 – 27 | 3.7 (3.5) | 11.4 (5.6) | 100 | $t$ = -14.77 | <.001^ |
| 2 | Burnout score | 1 – 5 | 2.3 (0.8) | 3.4 (1.0) | 51 | $t$ = -11.50 | <.001^ |
| 3 | Self-blame—Brief COPE Score | 2 – 8 | 2.8 (1.2) | 4.3 (1.7) | 48 | t = -8.88 | <.001^ |
| 4 | Fear to contract COVID-19 | 0 – 4 | 1.9 (1.2) | 2.9 (0.9) | 47 | $t$ = -9.47 | <.001^ |
| 5 | Concern to spread COVID-19 | 0 – 4 | 2.4 (1.2) | 3.4 (0.9) | 42 | $t$ = -9.77 | <.001^ |
| 6 | Intend to leave healthcare field | 0 – 4 | 0.4 (0.8) | 1.2 (1.2) | 34 | $t$ = -7.15 | <.001^ |

| Rank | Variable names | Scale Range | Low risk group (N=309) | High risk group (N=137) | VIP | Statistical analysis | *p*-value |
|---|---|---|---|---|---|---|---|
| 7 | Job difficulty | 0 – 4 | 2.1 (1.0) | 2.9 (1.0) | 27 | *t* = -7.92 | <.001^ |
| 8 | Organizational support | 0 – 24 | 16.1 (6.9) | 10.8 (8.1) | 23 | *t* = 6.58 | <.001^ |
| 9 | Intend to switch units / teams | 0 – 4 | 0.5 (1.0) | 1.4 (1.4) | 20 | *t* = -5.98 | <.001^ |
| 10 | Feel stigmatized | 0 – 4 | 0.9 (1.0) | 1.5 (1.4) | 19 | *t* = -5.14 | <.001^ |
| 11 | Provided psychosocial care training (employer) | Yes | 120 (38.8) | 30 (21.9) | 18 | $\chi^2$ = 14.81 | <.001* |
| | | No | 90 (29.1) | 61 (44.5) | | | |
| | | Not sure | 99 (32.0) | 46 (33.6) | | | |
| 12 | Behavioral disengagement—Brief COPE Score | 2 – 8 | 2.5 (0.9) | 3.4 (1.6) | 17 | *t* = -6.30 | <.001^ |
| 13 | Years of practicing | - | 20.8 (10.7) | 20.4 (10.6) | 15 | *t* = 0.33 | 0.74^ |
| 14 | Venting—Brief COPE Score | 2 – 8 | 4.2 (1.6) | 5.0 (1.4) | 13 | *t* = -5.57 | <.001^ |
| 15 | Support from family | 1 – 7 | 6.0 (1.3) | 5.4 (1.6) | 12 | *t* = 3.74 | <.001^ |
| 16 | Insufficient resources | 1-5 | 2.3 (1.1) | 2.8 (1.2) | 11 | *t* = -4.80 | <.001^ |
| 17 | CD-RISC 10 Resilience score | 0 – 40 | 32.2 (5.0) | 30.0 (5.6) | 11 | *t* = 3.83 | <.001^ |
| 18 | Denial—Brief COPE score | 2 – 8 | 2.4 (0.8) | 2.9 (1.4) | 11 | *t* = -4.18 | <.001^ |
| 19 | Support from significant others | 1 – 7 | 6.2 (1.4) | 5.8 (1.6) | 11 | *t* = 2.13 | 0.034^ |
| 20 | Intend to leave employer | 0 – 4 | 0.8 (1.1) | 1.4 (1.4) | 10 | *t* = -4.89 | <.001^ |

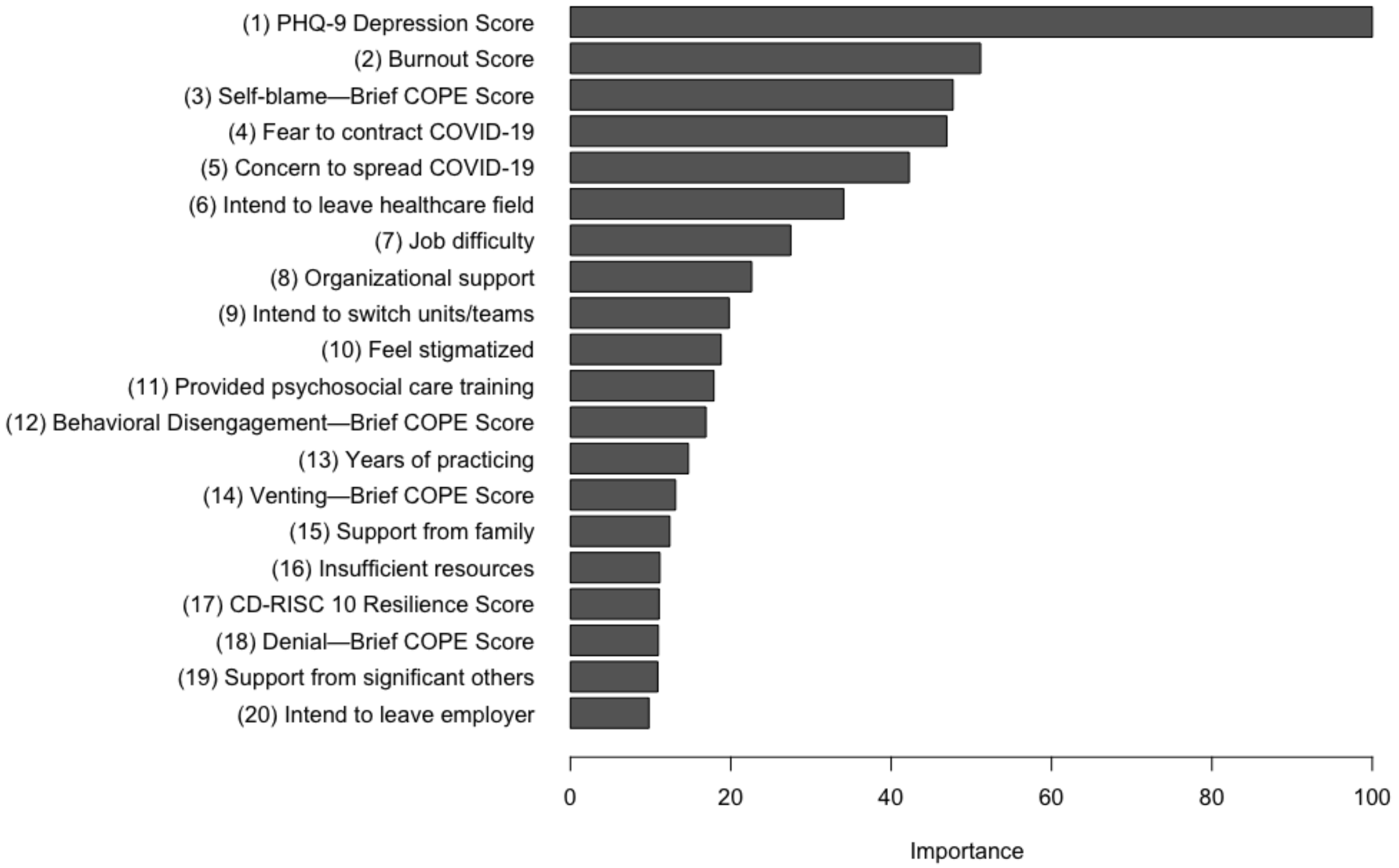

**Fig. 1.** Key predictors associated with PTSD risk in practicing physicians during COVID-19 pandemic

A cohort of **cognitive/psychological variables**—depression, burnout, fear—top the list of variables associated with higher risk of PTSD (**Fig. 1**). PTSD risk increased dramatically even when mild–moderate depressive symptoms were present (**Fig. 2-1**). PTSD risk became significantly prevalent with moderate–high burnout levels (burnout score≥3) (**Fig. 2-2**). Also, two types of fear—fear of contracting COVID-19 (**Fig. 2-4**), fear of transmitting it to loved ones (**Fig. 2-5**)—coincided with higher PTSD risk, presenting a "V-shaped," nonlinear relationship. Additionally, three **coping strategies** that were associated with increased PTSD risk include self-blame, venting, and behavioral disengagement. Self-blame (**Fig. 2-3**) and behavioral disengagement (**Fig. 2-12**) demonstrated a strong linear correlation with PTSD risk, whereas, venting presented a "V-shaped" curvilinear relationship with PTSD risk (**Fig. 2-14**). Three **occupational characteristics**—increases in job difficulty (**Fig. 2-7**), lack of resources (**Fig. 2-16**), perceived stigma for working with COVID-19 patients (**Fig. 2-10**)—were also associated with increases in PTSD risk. Physicians demonstrated resilience to all the challenges until they reached their highest levels. Among **demographics**, only age influenced PTSD risk, where service beyond 30 years demonstrated a positive association with higher risk of PTSD (**Fig. 2-13**). Lastly, **attrition variables**, represented by physician's intention to switch medical units (**Fig. 2-9**), leave their employer (**Fig. 2-20**), or leave healthcare entirely (**Fig. 2-6**), were positively associated with increased risk of PTSD. Our data show that frontline physicians with higher risk of PTSD have higher intentions to switch medical units (1 in 2.5[2]), leave current employer (1 in 2.2), or even leave the healthcare industry altogether (1 in 3.2) compared to their peers with lower risk of PTSD (switch teams=1 in 6.5; leave current employer=1 in 4.75; leave healthcare industry=1 in 10).

---

[2] "1 in *X*" can be interpreted as 1 out of *X* US physicians, who are working at the frontlines treating COVID-19 patients and have displayed higher levels of PTSD symptoms, have moderate to high likelihood of switching the teams or leaving their current employer or leaving the healthcare industry altogether in the next 2 years.

Numerous variables appeared to be **protective in nature**. The degree to which people felt supported by loved ones, such as friends/family (**Fig. 2-15**), significant others (**Fig. 2-19**) and their organization (**Fig. 2-8**) coincided with lower risk of PTSD (Blekas et al., 2020). PTSD risk spiked slightly as participants reported higher degrees of organizational support (**Fig. 2-8**). Participants who received training from employers in psychosocial care reported lower risk of PTSD **(Fig. 2-11)** compared to their peers who didn't receive any such care or were unsure of receiving such care. Increases in **resiliency** were associated with drops in PTSD risk (**Fig. 2-17**); however, PTSD risk begins to climb at the highest resilience levels. Lastly, **Fig. 2-18** suggests that as participants relied more on denial, PTSD risk was higher.

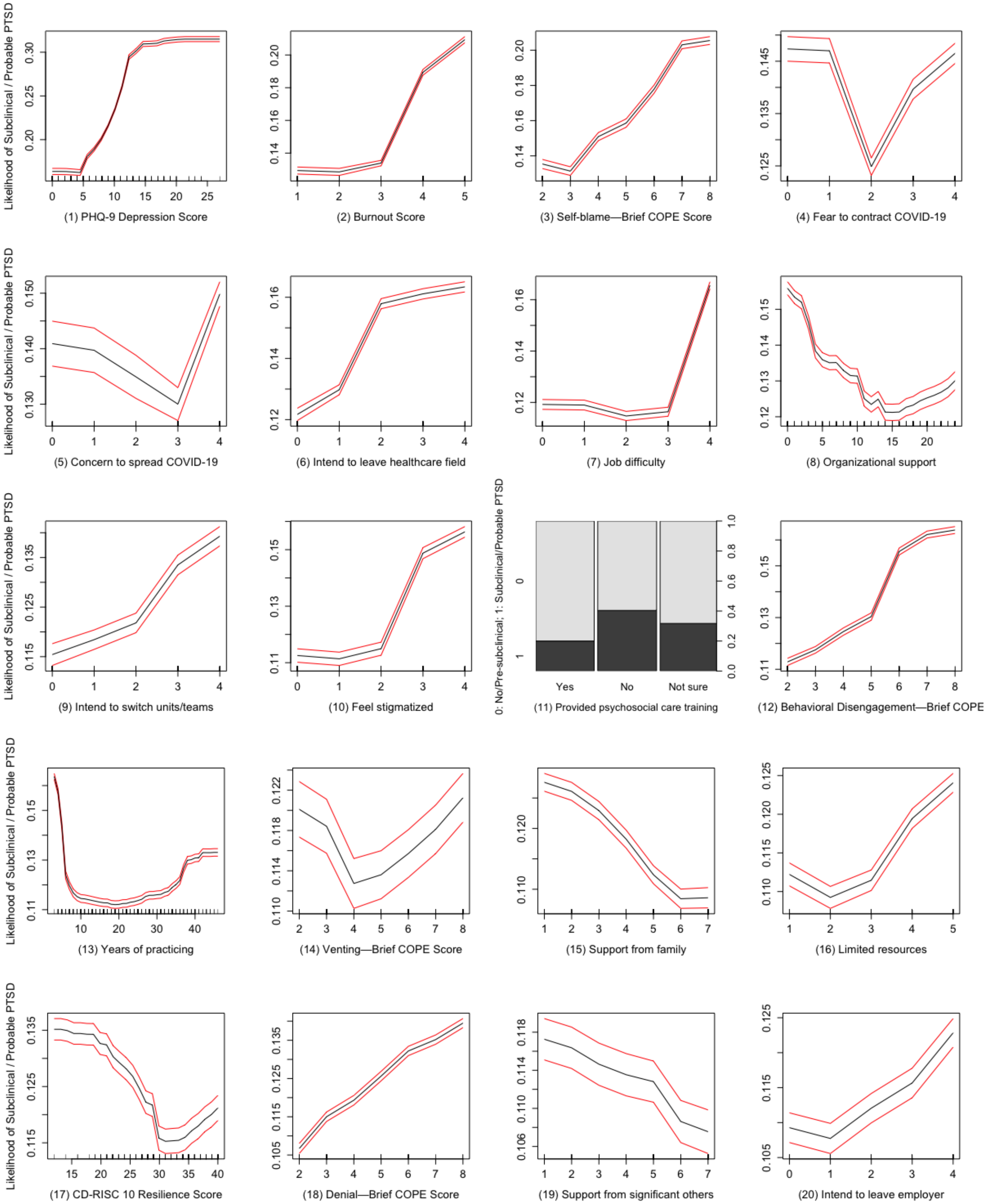

**Fig. 2**. Relationships of top 20 predictors to increase the likelihood of developing subclinical and probable PTSD symptoms (black curve is the average marginal effect of the predictor variable; red lines indicate the 95% confidence intervals)

## 4. Discussion

Physicians on the frontlines of COVID-19 are in crisis. In our study, among physicians across multiple specialties and states working directly with COVID-19 patients, 10.74% were classified as having probable PTSD and 20.08% as having sub-clinical PTSD, for a combined total of 30.82%. By comparison, in the US general population the lifetime risk of developing PTSD by age 75 is 8.7% and the twelve-month prevalence is 3.5% (American Psychiatric Association, 2013).

**Table 2** shows frontline physicians had significantly higher PCL-5 item scores and total scores when the full range of scoring was used. These results revealed the symptoms associated with an increased risk of PTSD for frontline physicians, as well as the need to identify the at-risk physicians for interventions. Our results also support the assumption that working with COVID-19 patients meets PTSD Criterion A of the DSM-5 (having experienced a traumatic event).

The findings in **Table 2** underscore not only how overall rates of PTSD differ between frontline and second-line physicians, but also how symptoms vary between these groups. The *International Classification of Diseases, 11th Edition* (World Health Organization, 2018) in aiming to create an abbreviated symptom assessment for PTSD, selected six items (among the 20 symptoms of PTSD included in DSM-5 and PCL-5) that are the most specific to PTSD (Stein et al., 2014). These six items include flashbacks and nightmares (Criterion B, Re-experiencing), avoiding memories and avoiding external reminders (Criterion C, Avoidance), and hypervigilance and exaggerated startle response (Criterion D, Hyperarousal). Our analysis indicated that all six of these items differed significantly between frontline and second-line physicians (**Table 2**). Interestingly, three items on which both the physician groups endorsed similar and considerable rates (negative beliefs, difficulty concentrating, and trouble sleeping), are highly correlated with other dysphoric conditions, such as depression and anxiety (Silverstein et al., 2020). Rather than being specific to PTSD, these three items may instead indicate concurrent stress-related symptoms among physicians, which warrant further examination.

**Table 3** presents a non-traditional way of examining PCL-5 scores by looking at the frequency of individuals who score items in the clinically significant range (≥2, moderate to severe). While PCL-5 mean score comparisons are typically used to analyze differences between groups using the full range of scoring (0 to 4), they do not reveal the number of participants who endorse a particular score or range of scores. We chose to examine the number (percent) of frontline and second-line physicians who endorsed symptoms in the range of 2 to 4, out of the total number of physicians in each group who participated in the survey and answered all the PCL-5 questions. Use of the frequency data provides additional information about the number of physicians who have a clinically significant PTSD symptom. For example, from Table 3, it can be inferred that 33.3% of the frontline physicians and 27.67% of the second-line physicians reported having moderate to severe problems with sleep (PCL-5; item # 20) which is not statistically significant at 95% confidence level, whereas 20.08% of frontline physicians and only 10.0% of second-line physicians endorsed disturbing memories (PCL-5; item # 1), which is statistically significant at 95% confidence level. Further exploration of frequency data could allow for the identification of the most relevant clinical symptoms associated with PTSD-related disturbances in physicians.

Of major consequence, the modeling analysis indicated that the cognitive outcomes such as depression and burnout were the greatest predictors of increased PTSD risk in the physicians working with COVID-19 patients. Previous studies have also shown that PTSD risk among physicians is positively associated with depression and burnout (Jackson et al., 2019; Restauri and Sheridan, 2020). Although burnout often coincides with depression among physicians (Jackson et al., 2019), it has been shown to have a different construct (Menon et al., 2020). Consistent with this, burnout and depression were independent predictors of PTSD risk in our analyses. In fact, depression, which has long been found to co-occur with PTSD in the aftermath of traumatic events (Gainer et al., 2021; Liebschutz et al., 2007; Pajonk et al., 2012; Shih et al., 2010), was the greatest predictor of higher risk of PTSD in physicians treating COVID-19 patients in the aftermath of the pandemic. Its significance underscores the need for physicians and healthcare administration to remain vigilant for indicators of depression, potentially engaging in active monitoring of its presence. Unlike with depression, PTSD risk was positively associated with only high burnout levels. In addition, PTSD risk levels were observed to be high when fears of contracting or transmitting COVID-19 were minimal, dipped when these fears were moderate, then dramatically spiked when the fears were great. Although moderate fear appears protective, low and high levels of fear may follow patterns similar to those discussed earlier, with high fear triggering trauma and low fear serving as a proxy for negative coping strategies such as denial.

Although the cross-sectional nature of the data prevents inferring causality, the higher intention of the frontline US physicians with higher PTSD risk to switch units, leave their employer, or leave healthcare completely compared to their peers with lower PTSD risk or even all the physicians in general (Sinsky et al., 2017) have important implications for the future of the physician workforce. With up to 50% of physicians already suffering from chronic stress and burnout entering the pandemic (Badahdah et al., 2020; Blekas et al., 2020; Elbay et al., 2020;

Mosheva et al., 2020; West et al., 2018; Yates, 2020), these new, trauma-related burdens for frontline physicians may herald an exodus from the already strained US healthcare system.

Despite their considerable predictive power and increasing popularity in public health (Lo-Ciganic et al., 2019; Mukherjee et al., 2021, 2020; Mukherjee and Wei, 2021; Parikh et al., 2019; Wei and Mukherjee, 2020), to our knowledge, this is the first study to leverage state-of-the-art, machine learning algorithms to predict and evaluate the factors associated with higher risk of PTSD in frontline physicians. Our results demonstrate the value of nonparametric, nonlinear machine learning algorithms to reveal complex relationships between predictor variables and PTSD risk, outperforming more conventional linear logistic regression in sophistication and precision (Liu and Salinas, 2017; Wshah et al., 2019). Our modeling approach not only revealed interplay between damaging and protective factors for PTSD risk, but also invites speculation on the nature of curvilinear relationships between the key factors and PTSD risk. For instance, physicians who vented (complained or processed trauma experiences with others) the least are the ones who suffered from higher PTSD risk, while those who vented a moderate amount presented lower PTSD risk. We speculate that minimal venting may forego the benefits of externally processing the traumatizing events and provoke trauma-induced symptoms (Bodie and Burleson, 2008; Mallinckrodt et al., 2012), whereas high levels of venting are a proxy for turmoil, a phenomenon found in other high stress-contexts (Coyne and DeLongis, 1986; Mallinckrodt et al., 2012). Similarly, increased institutional support largely decreased the risk of PTSD; however, our modeling reveals a spike in the PTSD risk at the highest degrees of organizational support. As with venting, the highest levels of organizational support may be provided for those, who are in the greatest need. Lastly, the spike in PTSD risk at the highest levels of resilience may come from physicians coping via a form of denial—another factor within the top 20 PTSD risk predictors. Understanding and exploring such nuances will further inform both theory and practice, helping support the frontline physicians through such crises.

Among demographics, only age was found to be a key predictor of PTSD risk. Previous research suggests younger physicians are less resilient to COVID-related trauma (Chong et al., 2004; Lai et al., 2020; Lancee et al., 2008; Sim et al., 2004; Su et al., 2007; Wu et al., 2009); however, the extant literature argues trauma is additive, suggesting older physicians would be more vulnerable following COVID-related trauma exposure (Ogle et al., 2014). Our findings offer support for both bodies of work, with the younger suffering the most, but PTSD risk markedly increasing among the older physicians.

However, there are certain limitations of our study. It is to be noted that the cross-sectional nature of these analyses prevents us from drawing conclusions about the causal relationship between predictor variables and the PTSD risk. In the future, longitudinal studies that bolster participant uptake are needed to confirm and expound upon these findings. Additionally, the study provoked a relatively low response rate, which is common for unincentivized, voluntary surveys.

In summary, our study identified some of the key factors associated with higher risk of PTSD among practicing physicians in the US due to the prolonged impact of the COVID-19 pandemic. Identifying key intervention variables can help stakeholders develop the means to proactively support individuals at higher PTSD risk. Thus, either through mitigating damaging variables and/or bolstering protective variables, this research provides both physicians and their institutions useful information to defend against the trauma-related threats to physicians resulting from the current and possible future epidemics/pandemics. The stability and future of the US healthcare system may depend on it.

## Conflict of Interest

The authors report no conflict of interest.

## Funding

The study was supported by the 2019-20 SUNY Research Seed Grant Program RFP #20-03-COVID, Proposal ID # COVID202060.

## Disclaimer

The ideas expressed in the article are those of the authors and may not be interpreted as AMA policy.

## Supplementary Online Content

## eMethods 1. Classification of PTSD symptoms

The different groups of PTSD risk were created by using a combination of the PCL-5 score and DSM criteria. This produced four levels of PTSD symptom severity: (1) "Probable PTSD," scoring 2 or higher in three or more DSM-5 categories and a total PCL-5 score >=33; (2) "Subclinical PTSD," scoring 2 or higher in two categories and a PCL-5 score >=12 and <=33 (three respondents had a score of 34 or 35, but in less than 4 categories); (3) "Pre-subclinical," scoring 2 or higher in none or one category and a PCL-5 score >=12 and <33; and (4) "No PTSD," a PCL-5 total score <12, irrespective of DSM-5 item scores.

## eMethods 2. Machine Learning Models used in the Study

### Introduction

In the study, we leveraged eight models to predict the likelihood of developing PTSD symptoms, including the traditionally-used parametric Logistic regression (Glonek and McCullagh, 1995; Hosmer Jr et al., 2013), nonparametric tree-based models (i.e., random forest (Breiman, 2001), bagging (James et al., 2014), gradient boosting method(Friedman, 2002), Bayesian additive regression trees (Chipman et al., 2008)), and black-box methods (i.e., Naïve Bayes (Rish, 2001), support vector machines (Gunn, 1998; Hearst et al., 1998; Steinwart and Christmann, 2008), neural networks (Nielsen, 2015)). Machine learning algorithms have been gaining more attention in the field of public health recently (Lo-Ciganic et al., 2019; Mooney and Pejaver, 2018). Compared with the conventional linear statistical models, the major advantages of applying machine learning models are: 1) the ability to capture the underlying interdependent and nonlinear relationships of the data (Beam and Kohane, 2018; Mooney and Pejaver, 2018); 2) capable of discover specific patterns and trends that could be unknown to humans, and producing the strong predictive ability (Char et al., 2018).

### Logistic regression

Logistic regression (Glonek and McCullagh, 1995; Hosmer Jr et al., 2013) is a parametric model that assumes a linear relationship between the feature vector $X_i^T = (x_{i1}, x_{i2}, \dots, x_{in})$ and the log-odds of the response variable $y$ being a success or a failure. In this study, the response variable is representative of the likelihood of developing the PTSD symptoms among physicians and can be treated as a binary categorical variable, i.e., $y \in \{+1, -1\}$. Let $\Pr(y_i = 1|X_i, \beta) = p_i$. And, the mathematical formula is given by

$$logit(p_i) = log\frac{p_i}{1 - p_i} = s_i = \beta_0 + \sum_{j=1}^{n} \beta_j x_{ij}.$$

Then,

$$p_i = sigmoid(s_i) = \frac{e^{s_i}}{1 + e^{s_i}} = \frac{1}{1 + e^{-s_i}},$$

where the sigmoid function is the inverse of the logit function.

### Random Forest

Random forest (Breiman, 2001) is a non-parametric ensemble tree-based model that aggregating $B$ bootstrapped decision trees denoted as $T_b$. By constructing a set of single decision trees, random forest takes the average of every output generated by each tree, which usually yields the better predictive accuracy than the single decision tree model. The formulation is denoted as

$$f(X) = \frac{1}{\mathrm{B}} \sum_{b=1}^{B} T_b(X).$$

Random forest randomly selects a subset of the predictor variables that can potentially avoid the issue of overfitting the data, but this model is sensitive to outliers.

**Bagging**

Bootstrap aggregation or Bagging (James et al., 2014) for short is another ensemble method that used decision tree-based classifiers to reduce the variance. $B$ different bootstrapped training data are generated by sampling from the training data with replacement. Each of the training set is used to develop the single classification tree, and in total $B$ different classifiers are constructed. Each constructed tree is considered weak learner, and is independent of each other. Then, the test data is used to validate each single tree model among $B$ trees, and the final output is determined by the majority vote across all $B$ trees for the classification problem. The major advantage of Bagging is to reduce the variance and it can provide an unbiased estimate of the test error (a.k.a. out-of-bag error).

**Gradient Boosting Method (GBM)**

Boosting is also the ensemble technique that the predictors are made sequentially, and are dependent upon previous trees, whereas a random forest is a collection of decision trees that are built in parallel. Gradient boosting method (Friedman, 2002) is another popular machine learning technique for both regression and classification problems. In GBM, a loss function is typically defined as squared error in the regression problem and logarithmic loss in the classification problem. Each decision tree is considered as weak learner that fitted to a smaller number of terminal nodes. The first tree model is fitted using the small data, and the second tree is developed using the residuals from the first tree that attempts to reduce the error. This process is being implemented through a sequence of trees, and during each iteration, gradient descent is used to minimize the loss function when sequentially adding more trees.

### Bayesian Additive Regression Trees

Bayesian additive regression tree (BART) (Chipman et al., 2008) is a non-parametric, tree-ensemble method, where a set of single tree models are aggregated to approximate the outcome variable denoted by $Y$. Generally speaking, the BART model has two main parts: a sum-of-tree model and a regularization prior. The single regression tree $T$ is constructed by partitioning the space of a set of predictors into different sub-regions that are representative of nodes, where the decision rules (in this study, the classification rules) are determined by the sequence of splits from root node to leaf node. Mathematically, the BART can be given by the sum of $m$ regression trees, that is,

$$G(X) = \sum_{j=1}^{m} g\left(X; T_j, M_j\right),$$

where $g\left(X; T_j, M_j\right)$ assigns the parameters $\mu_{ij} \in M_j$ of the regression tree $T_j$ to the predictor $X$. We applied probit function to map a set of predictor variables into a binary probability denoting the likelihood of developing PTSD symptoms among healthcare physicians. The regularization prior term is leveraged to control the complexity of tree's structure, balancing between computationally cost and the accuracy of the function approximation.

### Naïve Bayes

Naïve Bayes (Rish, 2001) is grounded in the Bayes' theorem with an assumption of independence among a set of predictors. Bayesian classification provides prior knowledge in probabilistic model to capture uncertainty about the model in predicting the outcome response. In the classification problem, we aim to estimate the function that maps from $X$ to $Y$, or equivalently $P(Y|X)$. By applying the Bayes rule, $P(Y = y_i|X)$ can be given by

$$P(Y = y_i|X = x_k) = \frac{P(X = x_k|Y = y_i)P(Y = y_i)}{\sum_j \left(X = x_k\middle|Y = y_j\right)P\left(Y = y_j\right)}.$$

Where $y_m$ indicates the m-th possible class for outcome variable $Y$, and $x_k$ is the k-th predictor variable. The final response label is denoted as

$$Y \leftarrow \underset{y_k}{\operatorname{argmax}} P(Y = y_k) \prod_i P(X_i|Y = y_k).$$

**Support Vector Machines (SVM)**

Support vector machine (SVM) (Gunn, 1998; Hearst et al., 1998; Steinwart and Christmann, 2008) is a generalization the maximal margin classifier which essentially maximizes the soft margin. It essentially extends the support vector classifier by using a larger feature space using higher order polynomials of the predictors. To introduce non-linearity a smarter way is to use kernels. SVM is known to be robust to the outliners and computationally efficient when the number of feature dimensions is greater than the number of total data points.

**Neural Networks**

Neural networks (Nielsen, 2015) are the widely-used machine learning models that basically simulate the nature of the biological neurons. Typically, it has three common layers for data processing: input layer, hidden layer and output layer. Data are first passed through the input layer. Hidden layer is used for intermediate processing, and translates the weights from the input layer to the next layer (it could be either another hidden layer or output layer). Using an activation function to map the information from the hidden layer to the final output layer, the predictive values could be obtained. The activation function is used to incorporate the non-linearity of the data. We used feedforward neural network to develop models for the training set.

**Model Selection Process**

To achieve optimal generalization performance and select the most robust predictive model (i.e., the model which accurately predicts without overfitting the data), the complexity of a statistical model should be controlled through bias-variance trade-off using the most widely used cross validation technique (Hastie et al., 2008; James et al., 2013). More specifically, predictive accuracy of each model was calculated by implementing 30-fold[3] random hold-out validation tests where in each iteration, 20% of the data was randomly held out and the model was trained with the remaining data and tested using the held-out sample (Hastie et al., 2008; James et al., 2013). To evaluate the performance of the classification models, we leveraged the widely-used statistical metrics based on the Confusion matrix, such as F1 score, Area Under the Curve (AUC, or C-statistic), Accuracy (%), Recall (%) and Precision (%), for both goodness-of-fit and predictive accuracy (Forman and Scholz, 2010). F1 score computed by the harmonic mean of Precision and Recall was applied to balance between positive predictive value and the true positive rate. AUC was calculated by the area under the receiver operating characteristic (ROC) curve. F1 score and AUC, which are mainly used in unbalanced datasets, were leveraged in this study to evaluate models' performance (Forman and Scholz, 2010). Finally, the model that outperformed all the other models in terms of goodness-of-fit and predictive accuracy was selected as the final predictive model for statistical inferencing. This is a well-known statistical process that ensures that the machine learning models do not overfit the data and provide accurate predictive performance, while providing interpretability benefits (Hastie et al., 2008; James et al., 2013).

[3] Given that cross-validation is based on randomized sub-sectioning of the data into 80–20% portions, 30 times repetition is naturally a conservative measure to ensure all the values have been used at least once.(Hastie et al., 2008; James et al., 2013)

## eMethods 3. Statistical Inference: Variable Importance (VI) and Partial Dependence Plots (PDP)

Although the non-parametric models outperform parametric models in terms of predictive performance, the improved predictability comes at the cost of reduced interpretability. However, statistical inferencing can be conducted for the non-parametric models using the variable importance ranking and partial dependence plots (PDPs) (Hastie et al., 2008; James et al., 2013). The importance of the variables is depicted by the inclusion proportion of the variables which denote the number of times a particular variable has been selected to develop the model. To understand how a particular predictor variable, affect the response variable, the PDPs are used. The PDP is estimated as follows:

$$p_j(x_j) = \frac{1}{n}\sum_{i=1}^{n} p_j(x_j, x_{-j}, i)$$

Here, $p$ is the statistical response surface; $n$ denotes the number of observations, $x_{-j}$ represents all the independent variables except $x_j$.

**eTable 1.** Description of Variables

| Rank | Variable Names | Variable Description | Score range [min-max] / Levels | Low risk group (N=309) | High risk group (N=137) | P-value |
|---|---|---|---|---|---|---|
| 1 | PHQ-9 Depression Score | PHQ-9 Depression Score | 0-27 | 3.7 (3.5) | 11.4 (5.6) | <0.001 |
| 2 | Burnout Score | Burnout Score | 1-5 | 2.3 (0.8) | 3.4 (1.0) | <0.001 |
| 3 | Self-blame—Brief COPE Score | Self-blame—Brief COPE Score | 2-8 | 2.8 (1.2) | 4.3 (1.7) | <0.001 |
| 4 | Fear to contract COVID-19 | Extent of fear to contract COVID-19 due to working on the frontlines | 0-4 | 1.9 (1.2) | 2.9 (0.9) | <0.001 |
| 5 | Concern to spread COVID-19 | Extent of concern for transmitting COVID-19 to family, friends, or relatives | 0-4 | 2.4 (1.2) | 3.4 (0.9) | <0.001 |
| 6 | Intend to leave healthcare field | Likelihood to leave healthcare field entirely in the next 2 years | 0-4 | 0.4 (0.8) | 1.2 (1.2) | <0.001 |
| 7 | Job difficulty | Increase in job difficulty due to COVID-19 | 0-4 | 2.1 (1.0) | 2.9 (1.0) | <0.001 |
| 8 | Organizational support | Positive support from the employer / organization since the outbreak of COVID-19 pandemic | 0-24 | 16.1 (6.9) | 10.8 (8.1) | <0.001 |
| 9 | Intend to switch units/teams | Likelihood to switch units/teams in the next 2 years | 0-4 | 0.5 (1.0) | 1.4 (1.4) | <0.001 |
| 10 | Feel stigmatized | Extent of feeling stigmatized by others because of working with COVID-19 patients | 0-4 | 0.9 (1.0) | 1.5 (1.4) | <0.001 |
| 11 | Provided psychosocial care training | Whether employer offered online training programs to their employees on psychosocial care principles | Yes | 120 (38.8) | 30 (21.9) | 0.001 |
| | | | No | 90 (29.1) | 61 (44.5) | |
| | | | Not sure | 99 (32.0) | 46 (33.6) | |
| 12 | Behavioral Disengagement—Brief COPE Score | Behavioral Disengagement—Brief COPE Score | 2-8 | 2.5 (0.9) | 3.4 (1.6) | <0.001 |
| 13 | Years of practicing | Years of practicing medicine (including residency) | 3-46 | 20.8 (10.7) | 20.4 (10.6) | 0.74 |
| 14 | Venting—Brief COPE Score | Venting—Brief COPE Score | 2-8 | 4.2 (1.6) | 5.0 (1.4) | <0.001 |
| 15 | Support from family | Extent of getting emotional help and support from family | 1-7 | 6.0 (1.3) | 5.4 (1.6) | <0.001 |
| 16 | Insufficient resources | Likelihood of having sufficient resources to take care of all the COVID-19 patients | 1-5 | 2.3 (1.1) | 2.8 (1.2) | <0.001 |
| 17 | CD-RISC 10 Resilience Score | CD-RISC 10 Resilience Score | 0-40 | 32.2 (5.0) | 30.0 (5.6) | <0.001 |
| 18 | Denial—Brief COPE Score | Denial—Brief COPE Score | 2-8 | 2.4 (0.8) | 2.9 (1.4) | <0.001 |
| 19 | Support from significant others | Extent of receiving emotional help and support from significant others | 1-7 | 6.2 (1.4) | 5.8 (1.6) | 0.027 |
| 20 | Intend to leave employer | Likelihood to leave current employer in the next 2 years | 0-4 | 0.8 (1.1) | 1.4 (1.4) | <0.001 |
| 21 | Employer evenly distributes your workload | Employer adequately implemented shifting of tasks to evenly distribute your workload | Yes | 147 (47.6) | 45 (32.8) | 0.001 |
| | | | No | 76 (24.6) | 56 (40.9) | |
| | | | Not sure | 86 (27.8) | 36 (26.3) | |
| 22 | No support from the employer | No support from the employer / organization since the outbreak of COVID-19 pandemic | 0-24 | 9.2 (7.3) | 12.9 (7.9) | <0.001 |
| 23 | Age | Age | 29-75 | 50.7 (11.0) | 50.3 (11.1) | 0.701 |
| 24 | Use of Emotional Support—Brief COPE Score | Use of Emotional Support—Brief COPE Score | 2-8 | 5.4 (1.8) | 5.5 (1.8) | 0.457 |
| 25 | Substance Use—Brief COPE Score | Substance Use—Brief COPE Score | 2-8 | 2.6 (1.2) | 3.2 (1.8) | <0.001 |
| 26 | Planning—Brief COPE Score | Planning—Brief COPE Score | 2-8 | 5.2 (1.8) | 5.9 (1.8) | 0.001 |
| 27 | Concern while working with patients | Having underlying health condition(s) that was of concern while working with patients during the COVID-19 pandemic | Yes | 87 (28.2) | 60 (43.8) | 0.005 |
| | | | No | 213 (68.9) | 73 (53.3) | |
| | | | Not sure | 9 (2.9) | 4 (2.9) | |
| 28 | Working status | Working status | Full-time | 276 (89.3) | 115 (83.9) | 0.245 |
| | | | Part-time | 28 (9.1) | 16 (11.7) | |
| | | | Furloughed | 2 (0.6) | 3 (2.2) | |
| | | | Laid off | 0 (0.0) | 1 (0.7) | |
| | | | On leave | 3 (1.0) | 2 (1.5) | |
| 29 | Use of Instrumental Support—Brief COPE Score | Use of Instrumental Support—Brief COPE Score | 2-8 | 4.4 (1.7) | 4.6 (1.8) | 0.3 |
| 30 | Support from friends | Extent of receiving emotional help and support from friends | 1-7 | 5.9 (1.3) | 5.4 (1.6) | <0.001 |
| 31 | Exposed to aerosol generating procedures with COVID-19 patients | Exposed to aerosol generating procedures (e.g., Nebulizer, HFNC, NIPPV, etc.) with suspected or known COVID-19 patients | 0 | 146 (47.2) | 54 (39.4) | 0.152 |
| | | | >= 1 | 163 (52.8) | 83 (60.6) | |
| 32 | Assist in deciding to allocate ICU bed/ventilators | Frequency of being part of the decision-making process about allocating ICU bed and/or ventilators | 1(always)-5(never) | 4.4 (1.2) | 4.4 (1.2) | 0.795 |
| 33 | Primary work setting | Primary work setting | Academic medical center | 84 (27.2) | 24 (17.5) | 0.204 |
| | | | Group practice | 75 (24.3) | 37 (27.0) | |
| | | | Hospital | 81 (26.2) | 40 (29.2) | |
| | | | Solo practice | 28 (9.1) | 12 (8.8) | |
| | | | Two-physician practice | 12 (3.9) | 5 (3.6) | |
| | | | Outpatient center | 12 (3.9) | 12 (8.8) | |
| | | | Others | 17 (5.5) | 7 (5.1) | |
| 34 | Hours worked in an average week before COVID-19 | Hours worked in an average week before COVID-19 outbreak | 0-100 | 48.6 (14.5) | 49.3 (15.2) | 0.642 |
| 35 | Employer provides psychological support hotline | If the employer arranged for a psychological support hotline for the employees' wellbeing at the organization | Yes | 164 (53.1) | 57 (41.6) | 0.068 |
| | | | No | 73 (23.6) | 37 (27.0) | |
| | | | Not sure | 72 (23.3) | 43 (31.4) | |
| 36 | Self-Distraction—Brief COPE Score | Self-Distraction—Brief COPE Score | 2-8 | 4.9 (1.7) | 5.5 (1.6) | <0.001 |
| 37 | Felt mistreated due to race or ethnicity | Felt mistreated or stigmatized by patients and/or their family members because of race or ethnicity | No | 296 (95.8) | 125 (91.2) | 0.088 |
| | | | Yes | 13 (4.2) | 12 (8.8) | |
| 38 | Pregnant or have a new born during COVID-19 | If pregnant (self/partner) or had (have) a new born while working with patients during the COVID-19 pandemic | No | 287 (92.9) | 126 (92.0) | 0.887 |
| | | | Yes | 22 (7.1) | 11 (8.0) | |
| 39 | | | No | 211 (68.3) | 78 (56.9) | 0.027 |
| | Extra hours worked during COVID-19 | If worked extra hours to care for COVID-19 patients when the practice received largest volume of COVID-19 patients | Yes | 98 (31.7) | 59 (43.1) | |
| 40 | Religion—Brief COPE Score | Religion—Brief COPE Score | 2-8 | 4.2 (2.1) | 4.3 (2.1) | 0.767 |
| 41 | Active Coping—Brief COPE Score | Active Coping—Brief COPE Score | 2-8 | 5.1 (1.7) | 5.4 (1.8) | 0.148 |
| 42 | Acceptance—Brief COPE Score | Acceptance—Brief COPE Score | 2-8 | 6.6 (1.4) | 6.4 (1.4) | 0.333 |
| 43 | Race | Race | American Indian or Alaska Native | 1 (0.3) | 0 (0.0) | 0.477 |
| | | | Asian | 46 (14.9) | 16 (11.7) | |
| | | | Black or African American | 12 (3.9) | 5 (3.6) | |
| | | | Native Hawaiian or Pacific Islander | 0 (0.0) | 1 (0.7) | |
| | | | White | 222 (71.8) | 106 (77.4) | |
| | | | Others | 28 (9.1) | 9 (6.6) | |
| 44 | Immigrant Status | Immigrant Status | No | 265 (85.8) | 122 (89.1) | 0.427 |
| | | | Yes | 44 (14.2) | 15 (10.9) | |
| 45 | Living arrangements changes during COVID-19 | Made any changes to living arrangements (even if temporarily) during the pandemic, due to concerns of transmitting COVID-19 to family / others | No | 188 (60.8) | 53 (38.7) | <0.001 |
| | | | Yes | 121 (39.2) | 84 (61.3) | |
| 46 | Positive Reframing—Brief COPE Score | Positive Reframing—Brief COPE Score | 2-8 | 4.8 (1.7) | 4.7 (1.7) | 0.616 |
| 47 | Relationship status | Relationship status | Single | 49 (15.9) | 18 (13.1) | 0.049 |
| | | | Married | 240 (77.7) | 102 (74.5) | |
| | | | Partnered | 17 (5.5) | 17 (12.4) | |
| | | | Widow / Widower | 3 (1.0) | 0 (0.0) | |
| 48 | Ethnicity | Ethnicity | No | 273 (88.3) | 128 (93.4) | 0.141 |
| | | | Yes | 36 (11.7) | 9 (6.6) | |
| 49 | Sex / Gender | Sex / Gender | Male | 151 (48.9) | 61 (44.5) | 0.521 |
| | | | Female | 156 (50.5) | 74 (54.0) | |
| | | | Prefer not to answer | 2 (0.6) | 2 (1.5) | |
| 50 | Exposed to COVID-19 patients during intubation | Exposed to suspected or known COVID-19 patients during the process of intubation | 0 | 207 (67.0) | 82 (59.9) | 0.178 |
| | | | >= 1 | 102 (33.0) | 55 (40.1) | |
| 51 | Number of months worked with COVID patients | Number of months worked with COVID patients | 1-10 | 4.6 (1.5) | 4.8 (1.6) | 0.23 |
| 52 | Worked outside normal scope of clinical practice | If worked outside normal scope of clinical practice when the practice received highest volume of COVID-19 patients | No | 221 (71.5) | 85 (62.0) | 0.06 |
| | | | Yes | 88 (28.5) | 52 (38.0) | |
| 53 | Humor—Brief COPE Score | Humor—Brief COPE Score | 2-8 | 4.3 (1.8) | 4.3 (1.8) | 0.793 |

**eTable 2.** Comparison of Goodness-of-fit among a library of machine learning models

| | Logistic | Random Forest | Bagging | Naive Bayes | GBM | BART | SVM | NN |
|---|---|---|---|---|---|---|---|---|
| F1 Score | 0.94 | 1.00 | 1.00 | 0.87 | 1.00 | 0.93 | 0.94 | 0.98 |
| AUC | 0.89 | 1.00 | 1.00 | 0.81 | 1.00 | 0.85 | 0.86 | 0.96 |
| Accuracy, % | 91.27 | 100.00 | 99.80 | 82.62 | 100.00 | 89.21 | 90.65 | 97.03 |
| Recall, % | 95.17 | 100.00 | 99.97 | 85.92 | 100.00 | 95.96 | 98.24 | 98.37 |
| Precision, % | 92.48 | 100.00 | 99.75 | 88.71 | 100.00 | 89.31 | 89.35 | 97.41 |

Abbreviations: AUC, area under the curve; GMB, gradient boosting method; BART, Bayesian additive regression trees; NN, natural network.

**eTable 3:** Comparison of Prediction Performance among a library of machine learning models

| | Logistic | Random Forest | Bagging | Naive Bayes | GBM | BART | SVM | NN |
|---|---|---|---|---|---|---|---|---|
| F1 Score | 0.84 | 0.88 | 0.86 | 0.85 | 0.86 | 0.87 | 0.88 | 0.85 |
| AUC | 0.76 | 0.76 | 0.76 | 0.78 | 0.78 | 0.77 | 0.76 | 0.77 |
| Accuracy, % | 78.87 | 82.52 | 80.65 | 80.16 | 81.27 | 81.48 | 81.97 | 79.95 |
| Recall, % | 84.18 | 93.11 | 88.78 | 84.17 | 87.26 | 89.58 | 92.79 | 85.64 |
| Precision, % | 84.81 | 83.40 | 84.06 | 86.53 | 85.75 | 84.46 | 82.98 | 85.26 |

Abbreviations: AUC, area under the curve; GBM, gradient boosting method; BART, Bayesian additive regression trees; NN, neural network.

Based on the prediction performance, the tree-based models (interpretable machine learning models) such as random forest (F1=0.88; 95% CI, 0.87-0.89), bagging (F1= 0.86; 95% CI, 0.85-0.87), gradient boosting method (F1=0.86; 95% CI, 0.85-0.88) and Bayesian additive regression trees (F1=0.87; 95% CI, 0.86-0.88), outperformed the black-box algorithms (i.e., non-interpretable machine learning models) such as Naïve Bayes (F1=0.85; 95% CI, 0.84-0.87) and neural network (F1=0.85; 95% CI, 0.84-0.86). On the other hand, the traditionally used logistic regression performed worst with the least F1 score (F1=0.84; 95% CI, 0.83-0.86). Note, although the support vector machine algorithm demonstrated a superior predictive performance to the random forest in terms of F1 score (F1=0.88; 95% CI, 0.86-0.89), it was not selected because the superior performance was obtained at the cost of reduced interpretability. Moreover, in terms of AUC measure, random forest also exhibited a higher predictive power (AUC=0.76; 95% CI, 0.75-0.78) than the other models, although they were not statistically different based on the paired t-test (***eTable 5***).

**eTable 4.** Significant test of F1 Score in out-of-sample prediction performance

| | **Logistic** | **Random Forest** | **Bagging** | **Naive Bayes** | **GBM** | **BART** | **SVM** | **NN** |
|---|---|---|---|---|---|---|---|---|
| **Logistic** | 1.000 | 0.000 | 0.042 | 0.414 | 0.036 | 0.010 | 0.001 | 0.284 |
| **Random Forest** | 0.000 | 1.000 | 0.034 | 0.004 | 0.074 | 0.271 | 0.648 | 0.001 |
| **Bagging** | 0.042 | 0.034 | 1.000 | 0.252 | 0.841 | 0.397 | 0.113 | 0.221 |
| **Naive Bayes** | 0.414 | 0.004 | 0.252 | 1.000 | 0.210 | 0.076 | 0.016 | 0.893 |
| **GBM** | 0.036 | 0.074 | 0.841 | 0.210 | 1.000 | 0.538 | 0.194 | 0.183 |
| **BART** | 0.010 | 0.271 | 0.397 | 0.076 | 0.538 | 1.000 | 0.520 | 0.054 |
| **SVM** | 0.001 | 0.648 | 0.113 | 0.016 | 0.194 | 0.520 | 1.000 | 0.007 |
| **NN** | 0.284 | 0.001 | 0.221 | 0.893 | 0.183 | 0.054 | 0.007 | 1.000 |

Paired t-test is used to test the statistical significance in mean of F1 Score between any two models in out-of-sample prediction performance across 30 iterations. The null hypothesis is: there is no difference between in population mean of F1 Score. The p-values are shown in the eTable 3, and the significant difference is highlighted in red ($p<0.05$).

**eTable 5.** Significant test of AUC in out-of-sample prediction performance

| | Logistic | Random Forest | Bagging | Naive Bayes | GBM | BART | SVM | NN |
|---|---|---|---|---|---|---|---|---|
| **Logistic** | 1.000 | 0.505 | 0.772 | 0.076 | 0.071 | 0.224 | 0.957 | 0.370 |
| **Random Forest** | 0.505 | 1.000 | 0.699 | 0.188 | 0.182 | 0.488 | 0.526 | 0.827 |
| **Bagging** | 0.772 | 0.699 | 1.000 | 0.112 | 0.106 | 0.318 | 0.807 | 0.537 |
| **Naive Bayes** | 0.076 | 0.188 | 0.112 | 1.000 | 0.965 | 0.576 | 0.075 | 0.226 |
| **GBM** | 0.071 | 0.182 | 0.106 | 0.965 | 1.000 | 0.591 | 0.070 | 0.219 |
| **BART** | 0.224 | 0.488 | 0.318 | 0.576 | 0.591 | 1.000 | 0.229 | 0.581 |
| **SVM** | 0.957 | 0.526 | 0.807 | 0.075 | 0.070 | 0.229 | 1.000 | 0.382 |
| **NN** | 0.370 | 0.827 | 0.537 | 0.226 | 0.219 | 0.581 | 0.382 | 1.000 |

Paired t-test is used to test the statistical significance in mean of AUC between any two models in out-of-sample prediction performance across 30 iterations. The null hypothesis is: there is no difference between in population mean of AUC. The p-values are shown in the eTable 3, and there is no significant difference among models in terms of AUC.

**eFigure 1.** Flow of data preprocessing

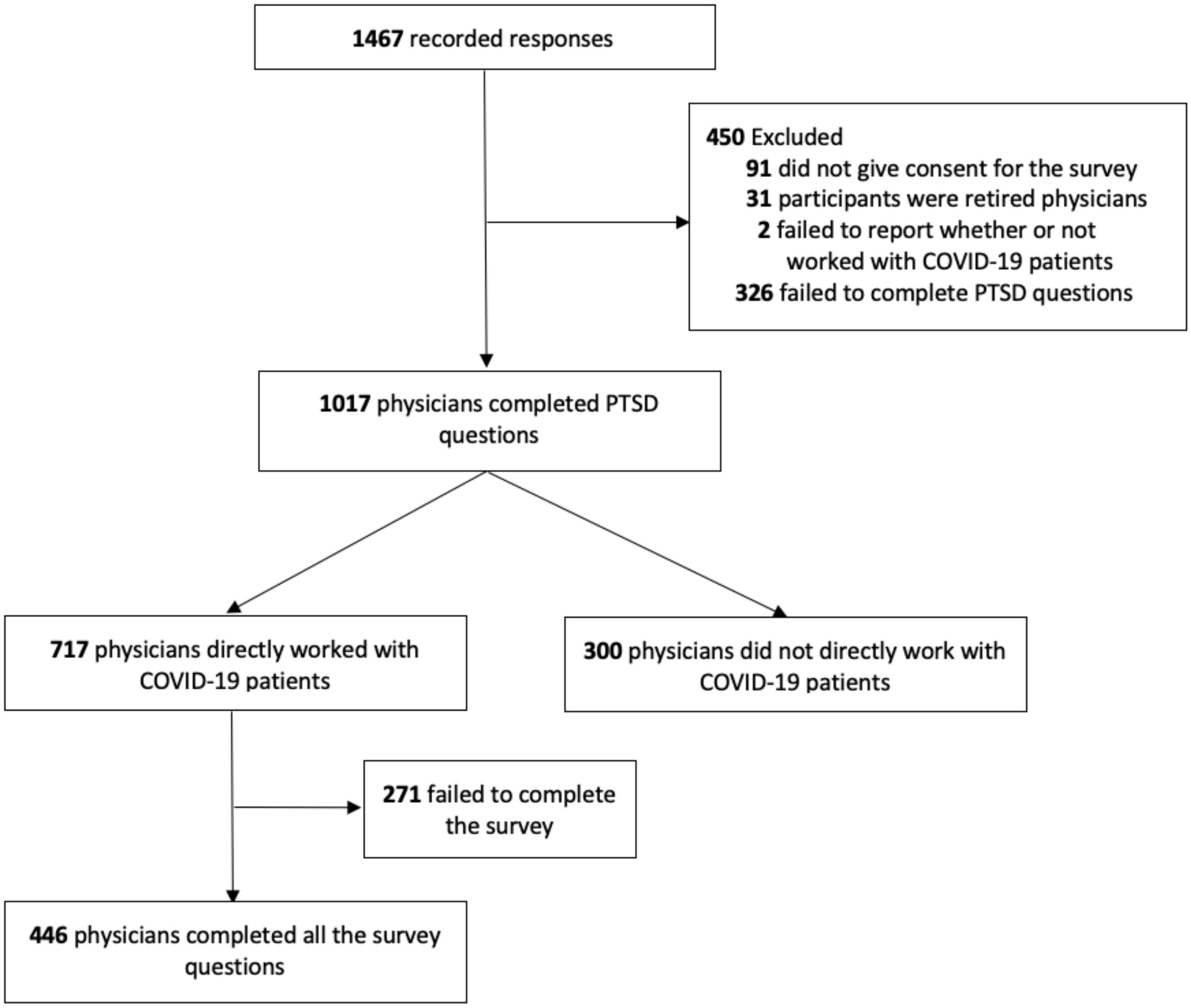

**eFigure 2.** Model performance (F1 Score and AUC) distribution in out-of-sample

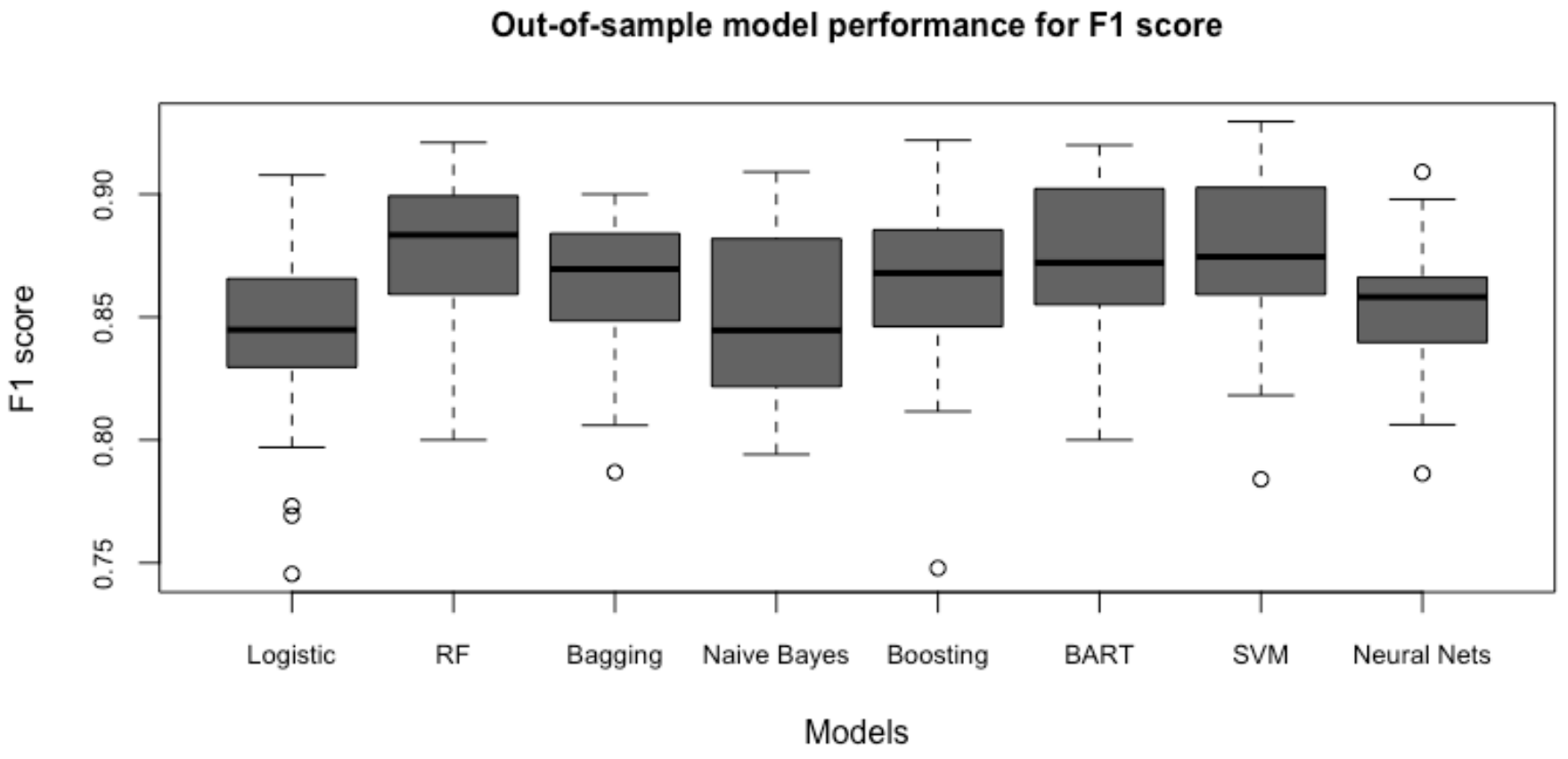

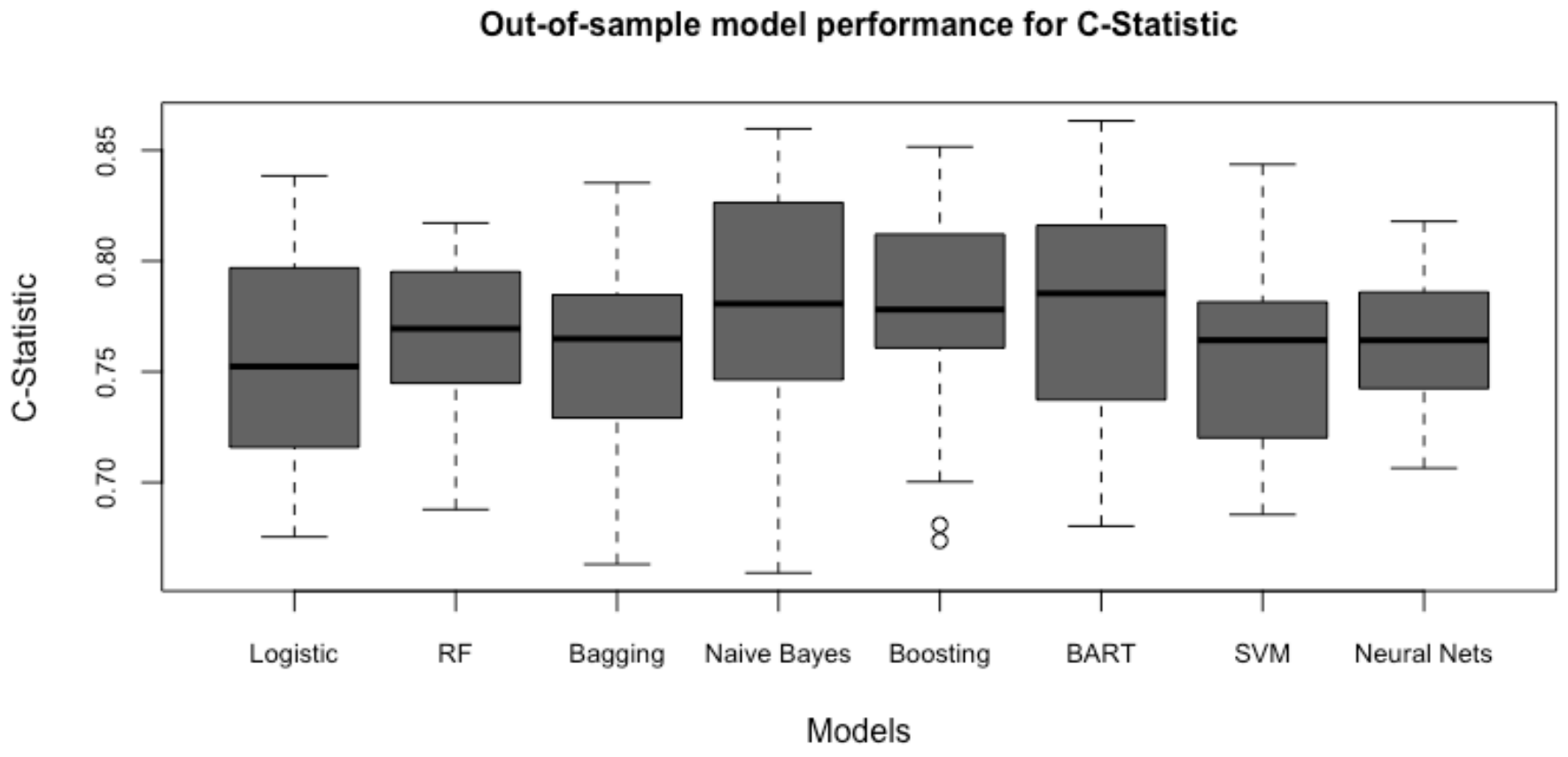

## eReferences